\documentclass[epj,nopacs]{svjour}
\usepackage{graphicx,verbatim,color} \usepackage{amsmath,amssymb,bm}
\usepackage{float}

\def\p{\partial} 
\def\eps{\varepsilon}
 
\newcommand{\be}{\begin{equation}} 
 
\newcommand{\ee}{\end{equation}} 
\newcommand{\bk}{{\bm k}}
\newcommand{\bx}{{\bm x}} 
 
\newcommand{\bz}{{\bm z}}
\newcommand{\bv}{{\bm v}} 
\newcommand{\bF}{{\bm f}} 
\newcommand{\bp}{{\bm p}}
\newcommand{\bq}{{\bm q}} 
\newcommand{\bu}{{\bm u}}
\newcommand{\buperp}{{\bm u}^{\rm 2D}} 
\newcommand{\fbuperp}{\hat{\bm u}^{\rm 2D}} 
\newcommand{\fbv}{\hat{\bm v}} 
\newcommand{\bomega}{{\bm \omega}}

\def\hm{{\bm h}^-_{\bm k}} 
\def\hp{{\bm h}^+_{\bm k}} 
\def\hpm{{\bm h}^\pm_{\bm k}}

\newcommand{\repl}[1]{{\color{black}{#1}}}
\begin{document}
\title{Non-universal behaviour of helical two-dimensional three-component
turbulence}

\author{Moritz Linkmann\inst{1,2,}\thanks{\email{moritz.linkmann@physik.uni-marburg.de}}, Michele Buzzicotti\inst{1} \and Luca Biferale\inst{1}%
}                     
\institute{
Department of Physics and INFN, University of Rome `Tor Vergata', Via della Ricerca Scientifica 1, 00133 Rome, Italy 
\and 
Fachbereich Physik, Philipps-Universit\"at Marburg, Renthof 6, 35032 Marburg, Germany
} 
\date{}

\authorrunning{M. Linkmann, M. Buzzicotti \& L. Biferale}
\titlerunning{Non-universal behaviour of helical 2D3C turbulence}

\abstract{ The dynamics of two-dimensional three-component (2D3C) flows is
relevant to describe the long-time evolution of strongly rotating flows and/or
of conducting fluids with a strong mean magnetic field. We show that in
the presence of a strong helical forcing, the out-of-plane component ceases to
behave as a passive advected quantity and develops a nontrivial dynamics which
deeply changes its large-scale properties. We show that a small-scale helicity
injection correlates the input on the 2D component with the one on 
the out-of-plane component. As a result, the third component 
develops \repl{a non-trivial energy transfer}.  The latter is  
mediated
by homochiral triads, confirming the strong 3D nature of the leading dynamical
interactions.  In conclusion, we show that  the out-of-plane component in a
2D3C flow enjoys strong non-universal properties as a function of the degree
of mirror symmetry of the small-scale forcing. }

\maketitle
\section{Introduction}
\label{sec:introduction}
Two-dimensional three-component (2D3C) flows are characterised by a velocity
field $\bu = (u_1,u_2,u_3)$ whose components only depend on two spatial
coordinates, e.g. $u_i=u_i(x,y;t)$ for $1\leqslant i \leqslant3$. 
\repl{Such a flow is relevant also for much more complicated systems whose
dynamics appear to be directly connected with this simplified 2D geometry, 
i.e.~three-dimensional turbulence under strong rotation
\cite{Cambon89,Waleffe93,Smith99,Chen05,Mininni09,Gallet15a,Alexakis15,BiferalePRX2016,galtier2014theory},
or conducting flows with strong background magnetic fields
\cite{Moffatt67,Alemany79,Zikanov98,Gallet15,Alexakis11,Bigot11}. In both the
aforementioned examples, the development of a 
backward energy cascade is
observed. A possible explanation for this behaviour arises from the decoupling
between the 2D3C manifold and the rest of the 3D domain
\cite{smith1999transfer}, with the consequent constraint for the energy to
follow channels living on the 2D3C plane.}
Furthermore, the 2D3C Navier-Stokes equations are identical to those for the
advection of a passive scalar by a two dimensional flow, because the
$z$-component, $u_3(x,y;t)$, is only advected by the 2D-component
$\buperp= (u_1,u_2,0)$. We define $\theta \equiv u_3$ such 
that the 2D3C-Navier-Stokes equations for incompressible flow read
\begin{align}
\label{eq:2d3c-nse}
\p_t \buperp &= -(\buperp \cdot \nabla) \buperp - \nabla P + \nu \Delta \buperp \ ,\nonumber \\  
\p_t \theta &= -(\buperp \cdot \nabla) \theta + \nu \Delta \theta  \ , 
\end{align}
where $P=P(x,y)$ is the pressure, $\nu$ the kinematic viscosity and $\nabla
\cdot \buperp = 0$. The density has been set to unity for convenience. 
A turbulent 2D3C flow can thus be expected to display a
split energy cascade: The energy of the 2D-component shows an inverse
cascade, while the energy of the passive third component develops 
a direct cascade according to classical results for passive scalar 
advection in 2D turbulence \cite{Falkovich01} and arguments based on triadic
dynamics \cite{Biferale17}. \\

\noindent
In a 2D3C flow, the vorticity of $\buperp$ points to the $z$-direction only, that is
$\nabla \times \buperp = \omega \hat \bz$, where $\bz$ denotes the unit vector
in $z$-direction. The latter has interesting consequences concerning the
dynamics of the inviscid invariants, which are the total 2D energy, $E^{\rm 2D}$,
the total energy of $\theta$, $E^\theta$, and the kinetic helicity $H$. The
latter is the normalised $L^2$-inner product of velocity and vorticity
\be
H = \langle  \bu \cdot \bomega \rangle 
  = \frac{1}{|V|} \int_V d\bx \ \bu \cdot \bomega \ , 
\ee
where $\bomega = \nabla \times \bu$ is the 3D vorticity. In 
the present case, $H$ depends on $\theta$ and $\omega$ only 
\cite{Moffatt14,Biferale17}  
\be
H = 2 \langle \theta \omega \rangle \ .
\ee
Hence, a nonzero helicity in a 2D3C flow implies a  correlation between the
out-of-plane scalar and the vorticity of the advecting 2D velocity field.  In
this setting, nonzero helicity implies that the out-of-plane component 
is in fact no longer passive as it is correlated to the vorticity of the 
2D velocity field \cite{Biferale17}. 
Such a case is of particular interest for rapidly rotating flows, where
it is known that the presence of helicity affects 
the cascade dynamics \cite{Mininni09a,Teitelbaum09,Mininni10a,Mininni10b}. 
In the presence of static forcing helicity can also be generated dynamically 
in rotating flows \cite{Dallas16}.
\\ 

\noindent
How to enforce a correlation of the out-of-plane component with the 
2D vorticity and its consequences are the focus of the present work. 
Our main findings are the
following.  (i) In the presence of a strong small-scale helical forcing, the
out-of-plane component develops a non-trivial inverse energy transfer, at
difference from what happens when mirror symmetry is respected. We interpret 
this as evidence of non-universality of the whole 2D3C dynamics, i.e. of its strong
sensitivity to the helical properties of the external forcing. (ii) In the
presence of the non-trivial inverse transfer also for the third component,
there is a strong asymmetry among the dynamics of homochiral and heterochiral
Fourier triads, signature of the 3D features of the underlying
dynamics \cite{Biferale17,Biferale12,Biferale13a}. 
\noindent
This paper is organised in the following way. In section \ref{sec:2D3C-flows}
we discuss how the passive scalar and the 2D-vorticity can be correlated
through a particular choice of forcing, that is, a force with maximal helicity.
The latter is investigated numerically in section \ref{sec:numerics}, where we
describe the generated database and the main results. We summarise and discuss
our results in section \ref{sec:conclusions}.

\section{The dynamics of 2D3C flows}
\label{sec:2D3C-flows}
\noindent
In order to create a connection between $\theta$ and $\omega$, without altering the equations 
of motion, we need to use an external force which is fully helical. 
A fully helical force is constructed through a projection 
operation in Fourier space applicable to any square-integrable 
vector field $\bv$, which we assume here to be solenoidal.
Owing to the latter, each Fourier mode $\fbv_\bk$ of $\bv$ 
has two degrees of freedom given by fully helical 
basis vectors \cite{Constantin88,Waleffe92}, such that   
\be
\fbv_\bk(t) = \fbv_\bk^+(t) + \fbv_\bk^-(t) = \hat v^+_\bk(t)\hp + \hat v^-_\bk(t)\hm \ ,
\ee
where $\hpm$ are normalised eigenvectors of the curl operator in 
Fourier space. That is, the Fourier modes of the vector field $\bv$ 
are described by two components satisfying
\be
i\bk \times \fbv_\bk^{s_k}= s_kk \fbv_\bk^{s_k} \ ,
\ee
and $s_k = \pm$. Since $\hpm$ are orthonormal, a projection operation onto 
helical subspaces defined as the span of either $\hp$ or $\hm$ can be carried out. 
Here, we apply the helical projection to a square-integrable external force $\bF$, which 
is assumed to be solenoidal, too, in order to comply with the incompressibility
constraint on the velocity field.
%
\repl{More precisely, we decompose $\hat\bF_\bk$ into its positively and 
negatively helical components $\hat \bF^+_\bk$ and $\hat \bF^-_\bk$, \
such that the helicity of the forcing can be adjusted through suitable
linear combinations of $\hat \bF^+_\bk$ and $\hat \bF^-_\bk$, }
leading to the following equations of motion in Fourier space 
\repl{  
\be
\partial_t \hat \omega_\bk = - \sum_{\bq=\bk-\bp} (i\bk \cdot \fbuperp_\bp)\hat \omega_\bq + \nu k^2 \hat \omega_\bk + k (\hat \bF^+_\bk - x\hat \bF^-_\bk)_z \ ,
\label{eq:omega-k}
\ee

\be
\partial_t \hat \theta_\bk = - \sum_{\bq=\bk-\bp} (i\bk \cdot \fbuperp_\bp)\hat \theta_\bq + \nu k^2 \hat \theta_\bk + (\hat \bF^+_\bk + x\hat \bF^-_\bk)_z \ ,
\label{eq:theta-k}
\ee
} where the evolution equation of the 2D-component has been written in vorticity
form in order to highlight the specific nature of the forcing.  \repl{The
factor $x$, $0 \le x \le 1$, is a coefficient to regulate the fraction of
helicity injected by the forcing. In particular, for $x=0$ the forcing is
fully helical while for $x=1$ we recover the non-helical forcing.} The
structure of the 2D3C Navier-Stokes equations remains unaltered, only the
properties of the forcing have been adjusted. As such,  the dynamical system
described by eqs.~\eqref{eq:omega-k}-\eqref{eq:theta-k} can be  realised in
the laboratory, if precise control over the helicity of the forcing can be
achieved.

\repl{In case of fully helical forcing, i.e.~for $x=0$ and hence  $\hat\bF = \hat\bF^+$, 
the evolution of the 2D vorticity and of the out-of-plane component or the velocity field
are forced in a correlated way. 
As a result,  Eq.~\eqref{eq:theta-k} 
does not correspond any more to the
evolution of a linear passive quantity, because of the link between
the injection and the advection velocity term.  
It is well known that
in the latter case the scalar field must not necessarily develop a forward
cascade even if advected by an incompressible velocity field \cite{celani2004active}.
Moreover, if $\hat \bF^+_\bk$ is acting only in a single shell at $k=k_f$,
then $\hat \omega_\bk/k_f$ and $\hat \theta_\bk$ obey the same equation
of motion.} In configuration space, single-shell helical forcing results in
$\nabla \times \bF^+ = \alpha \bF^+$, \repl{where $\alpha$ is a coefficient
inversely proportional to the characteristic length scale of the force} and
$\bF^+$ the inverse Fourier transform of $\hat \bF^+$. The specific nature of
the force then results in the following equation for the difference $\xi
\equiv \theta - \alpha \omega$
\be
\label{eq:evolution-xi}
\partial_t \xi =- (\buperp \cdot \nabla)\xi + \nu \Delta \xi = 0 \ , 
\ee
where  the forcing is absent from the equation of motion 
for $\xi$. Furthermore, the structure of the equation shows that $|\xi|^2$ is 
an inviscid invariant. Hence Eq.~\eqref{eq:evolution-xi} implies that 
$\xi$ will decay in time.   
The latter can be used to connect the energy spectra of $\theta$ and $\bu^{\rm 2D}$ 
\begin{align}
E^\theta(k,t)  &=  \frac{1}{2}\sum_{|{\bk}|=k} |\hat\theta_\bk(t)|^2 \ , \\
E^{2D}(k,t)  &=  \frac{1}{2}\sum_{|{\bk}|=k} |\fbuperp_\bk(t)|^2 \ ,
\end{align}
in order to predict the scaling of $E^\theta(k,t)$. According to 
eq.~\eqref{eq:evolution-xi}, we expect $\lim_{t \to \infty} \xi(\bx,t) = 0$ 
 and therefore 
$\hat \theta_\bk =  \hat \omega_\bk/k_f$ asymptotically in time, which results in 
\be
\label{eq:spectra-helforce-shell}
E^\theta(k,t) = \frac{1}{2k_f^2}\sum_{|\bk|=k}|\hat \omega_\bk(t)|^2 
            = \frac{k^2}{k_f^2}E^{\rm 2D}(k,t) \ . 
\ee 
In the inverse energy cascade regime, i.e. for $k<k_f$, $E^{\rm 2D}(k,t) \sim k^{-5/3}$ 
\cite{Kraichnan67,Lilly69,Sommeria86,Gotoh98,Paret98,Boffetta12}, 
resulting in $E^\theta(k,t) \sim k^{1/3}$. 
\noindent 

\repl{However, in the more generic case, when $\bF^+$ acts in a wavenumber band, the forcing is no longer absent from the evolution equation for $\xi$. 
That is, in the band-forced case the asymptotic spectral scaling of $E^\theta(k,t)$ can only be investigated numerically. In what follows such an investigation is carried out by means of several series of direct numerical simulations (DNS). Numerical results, reported in the next section, show a very similar behaviour for the single shell and the band-forced systems, confirming also in the latter case a non-universal nature for the dynamics of the passive scalar.}

\begin{table*}[h]
\begin{center}
\begin{tabular}{ccccccccccc}
  Run id & $N$ & $U$ & $\varepsilon$ \tiny{$[\times10^{-3}]$} & $\ell$ & $\nu$ & $F$ & \repl{$x$} & $k_f$ & $t/T_f$ & \# \\
   \hline
  Hel1 & 256 & 0.49 & 1.25 & 1.33 & $1.8 \times 10^{-13}$ & 0.55 & \repl{0} & 20 & 67 & 20 \\ 
NonHel1 & 256 & 0.42 & 1.27 & 1.25 & $1.8 \times 10^{-13}$ & 0.55 & \repl{0} & 20 & 54 & 15 \\ 
  Hel2 & 256 & 0.48 & 1.56 & 1.36 & $1.8 \times 10^{-14}$ & 0.88 & \repl{0} & 32 & 67  & 1 \\ 
  Hel3 & 256 & 0.48 & 0.99 & 1.32 & $1.8 \times 10^{-14}$ & 0.13 & \repl{0} & 16-32 & 67 & 1 \\ 
  Hel4 & 256 & 0.48 & 1.52 & 1.22 & $1.8 \times 10^{-14}$ & 0.42 & \repl{0} & 30-34 & 67 & 1 \\ 
  Hel5 & 512 & 0.48 & 1.69 & 0.97 & $1.5 \times 10^{-16}$ & 1.80 & \repl{0} & 64 & 69 & 1 \\  
  \repl{Hel6} & 2048 & 0.34 & 1.12 & 0.5 & $5.7 \times 10^{-20}$ & 20.0 & \repl{0} & 200 & 32 & 2 \\  
  \repl{Hel7} & 2048 & 1.5 & 25.0 & 0.8 & $5.7 \times 10^{-20}$ & 20.0 & \repl{0} & 180-200 & 35 & 2 \\  
  \repl{FracHel1} & 2048 & 0.35 & 1.13 & 0.55 & $5.7 \times 10^{-20}$ & 20.0 & \repl{0.2} & 200 & 32 & 1 \\  
  \repl{FracHel2} & 2048 & 0.38 & 1.15 & 0.58 & $5.7 \times 10^{-20}$ & 20.0 & \repl{0.5} & 200 & 32 & 1 \\  
  \repl{FracHel3} & 2048 & 0.5 & 1.22 & 0.6 & $5.7 \times 10^{-20}$ & 20.0 & \repl{1} & 200 & 32 & 1 \\  
  \hline
  \end{tabular}
\caption{
Specifications of the numerical simulations.
$N$ is the number of grid points in each Cartesian coordinate, 
$U=\sqrt{2E}$ the root-mean-square velocity,
$\varepsilon$ the dissipation rate,
$\ell=(\pi/2U^2)\int dk \ E(k)/k$ the integral scale,
$\nu$ the kinematic hyperviscosity,
$F$ the magnitude of the forcing,
$k_f$ the wavenumber at which the forcing is active, 
$t/T_f$ the run time in units of
forcing-scale eddy turnover time
$T_f=(2\pi/(F k_f))^{1/2}$, and $\#$ the ensemble size.
The values for $U$, $\ell$ and $\eps$ 
are taken at the end of the simulations.
}
\end{center}
\label{tbl:simulations}
\end{table*}

\begin{figure}[tbp]
  \includegraphics[width=\columnwidth]{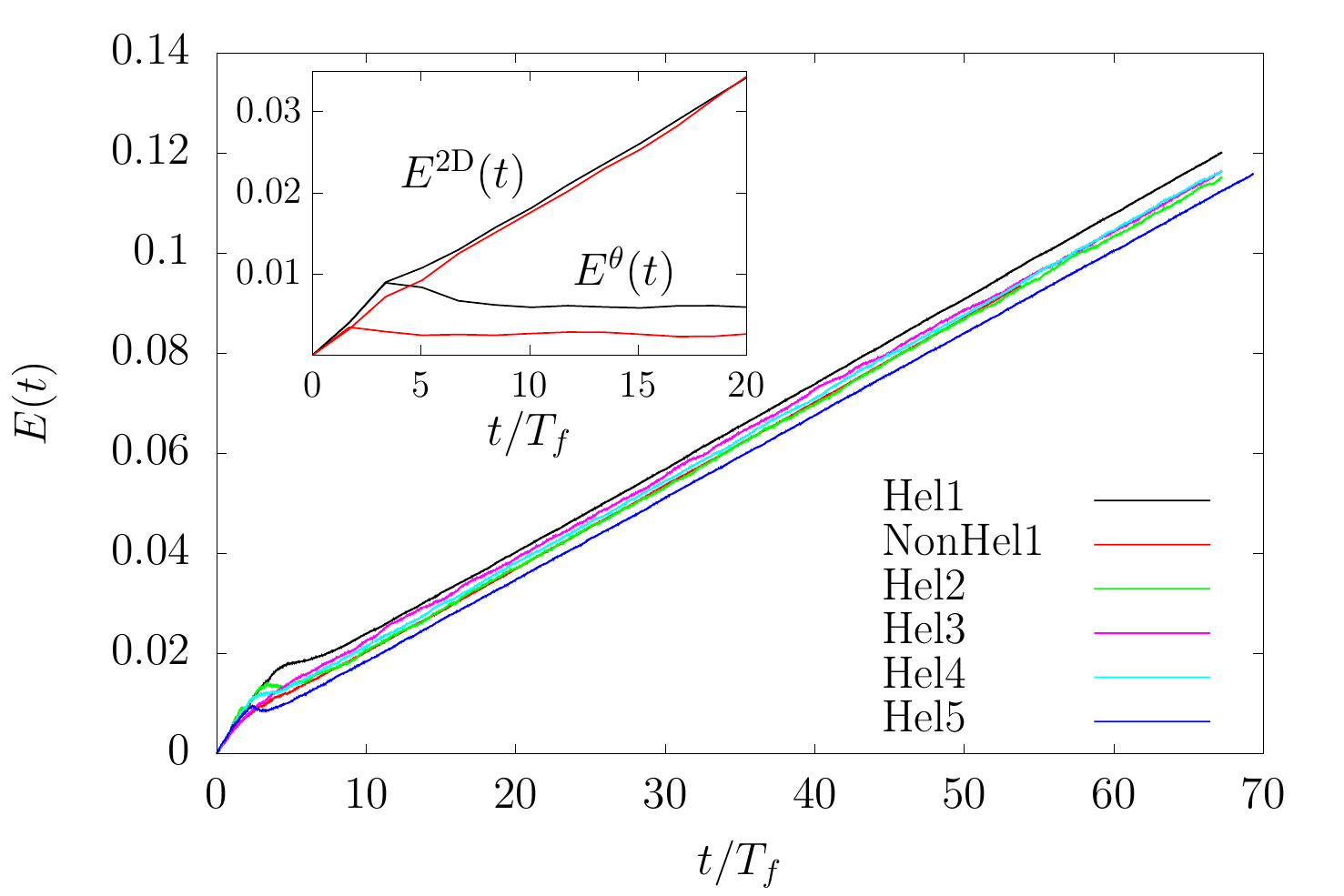}
\caption{
Time evolution of the total energy for datasets with either 
fully helical or nonhelical forcing.
Inset: Time evolution of $E^\theta(t)$ and $E^{\rm 2D}(t)$ up to $t/T_f=20$ for one realisation of 
datasets (Hel1) and (NonHel1), respectively.
The color scheme of the inset is the same as that of the main graph.
}
\label{fig:time-evol}       
\end{figure}

\section{Numerical simulations}
\label{sec:numerics}
We study numerically the dynamics of incompressible 2D3C flows given by eqs.~\eqref{eq:2d3c-nse}
subject to helical and
nonhelical forcing in order to shed light on the effect of the
forcing-induced correlation between $\theta$ and $\omega$ on the energy 
flux from small to large scales. Furthermore, we investigate the influence
of forcing on a bandwidth on the scaling properties of $E^\theta(k)$. 
For this purpose, it is necessary to inject energy into the system 
at the small scales, to ensure large scale separation between the forcing 
scale and the largest resolved scale in order to study the inverse cascade, 
while still resolving small-scale dynamics. 
We consider the 
Navier-Stokes equations with the Laplace operator replaced by 
higher-order hyperviscous dissipation 
\begin{align}
\label{eq:nse}
&\partial_t \bu = - \nabla \cdot (\bu \otimes \bu) - \nabla P + \nu (-1)^{n + 1}\Delta^{n}\bu + \bF \ , \\
\label{eq:incomp}
&\nabla \cdot \bu = 0 \ ,
\end{align}
where $n=4$. Equations \eqref{eq:nse}-\eqref{eq:incomp}
are solved numerically on a triply periodic domain $V=[0,2\pi]^3$ using a 
pseudospectral code with full dealiasing by truncation following the two-thirds rule. 
The forcing is given by a white-in-time random process in Fourier space
\be
\langle \hat \bF_\bk(t) \hat \bF^*_\bq(t') \rangle = F^2 \delta_{\bk,\bq} \delta(t-t')
\hat Q_\bk,
\ee
where $\hat Q_\bk$ is a projector applied to guarantee incompressibility
and $F$ is nonzero in a given band of Fourier modes concentrated at intermediate 
to small scales, see table 1 for details. 
Since the 2D component of u displays an inverse energy transfer and no
large-scale energy removal is used, the simulations will reach a statistically
stationary state only after very long evolution times. Here, we terminate the
simulations before the formation of a condensate at the largest resolved
scales.
In order to obtain statistical measurements 
we therefore generate ensembles over independent runs.
The simulations differ in the level of helicity injection, in the scale separation 
between the forcing scale and the largest resolved scale and in the width of the 
forcing band. Runs with fully helical forcing are labelled 
(Hel) followed by a number while those
carried out with helically neutral forcing are labelled (NonHel)
\repl{and those with fractionally helical forcing are labelled (FracHel).}
Further details on runtime, ensemble size, number of grid points, 
location of the forcing band, etc., are given in table 1. \\ 

\noindent
In order to compare between different datasets, the parameters are chosen 
such that the growth rate of the total energy (per unit volume), $dE/dt$, remains the same 
between different datasets. The energy is given as
\be
E(t) = \sum_{k\neq 0} E(k,t) \ ,
\ee  
where $E(k,t) = E^{\rm 2D}(k,t) + E^\theta(k,t)$ is the total energy spectrum. 
The energy of the 2D-component, $E^{\rm 2D}(t)$, and that of the out-of-plane component,
$E^\theta(t)$, are defined analogously.
The time evolution of $E(t)$ for all datasets is shown
in the main graph of Fig.~\ref{fig:time-evol}. The time evolution of 
$E^{\rm 2D}(t)$ and $E^\theta(t)$ for dataset (Hel1) and dataset (NonHel1) 
is presented in the inset of  
Fig.~\ref{fig:time-evol}, where one can see that the energy growth is 
due to the 2D component only, while the $\theta$-component reaches a 
statistically stationary state, with 
$E^\theta(t)$ being larger for dataset (Hel1) compared to dataset (NonHel1). 
We will come back to this point in section \ref{sec:fluxes}.
\\

\noindent
In what follows we distinguish between two main questions. 
Firstly, for helical forces we investigate the influence of extending the 
domain where the forcing is active from a single shell to a wider band
of Fourier modes. 
Second, we compare the effect of helical and nonhelical forces both active 
in a wavenumber shell centered at $k_f$ in order to investigate the effect of 
forcing-induced correlation between $\theta$ and $\omega$ on the 
dynamics. The main observables will be energy fluxes and spectra. 

\begin{figure*}[h]
\begin{center}
\includegraphics[width=\columnwidth]{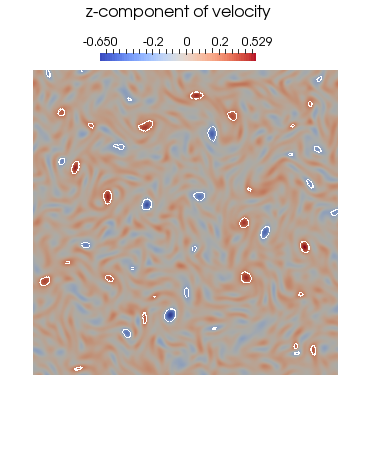}
\includegraphics[width=\columnwidth]{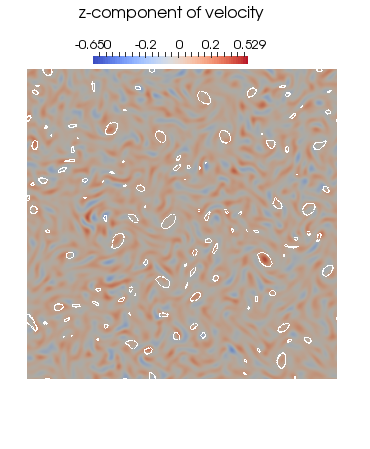}
\end{center}
\vspace{-2.2cm}
\begin{center} 
{\bf helical \hspace{8cm} nonhelical}
\end{center}
\caption{Visualisations of $\theta$ snapshots for datasets (Hel1) (left) and (NonHel1) (right) at $t/T_f = 25$,
the white contour lines indicate the regions with vorticity $|\omega| = 0.5$.
}
\label{fig:visualisations-T60}
\end{figure*}

\noindent
In order to clearly distinguish between the dynamics of helically and
nonhelically forced 2D3C flows, we generated two ensembles of runs from random
initial data with identical parameters concerning resolution, forcing band,
forcing magnitude and run time. The two ensembles only differed with respect to
the helicity of the forcing. 
In both cases the forcing is active on a single
wavenumber shell. According to the discussion in section
\ref{sec:2D3C-flows}, it can be expected that $\theta$ and $\omega/k_f$ behave
similarly for dataset (Hel1).
In contrast, no connection between $\theta$ and
$\omega/k_f$ is given through the equations of motion for dataset (NonHel1).
Visualisations of $\theta$ compared with contour lines of $|\omega|$ are shown
in Fig.~\ref{fig:visualisations-T60} for single realisations taken from
datasets (Hel1) (left panel) and (NonHel1) (right panel) at $t/T_f = 25$.  
As can be seen from the figures, intense regions of $\theta$
indeed coincide clearly with intense regions of $|\omega|$ for dataset (Hel1),
while no such effect can be identified for dataset (NonHel1). 
Visualisations of rapidly rotating flows subject to helical forcing show 
similar correlations between vorticity and the velocity component parallel
to the rotation axis \cite{Mininni10b}. \\ 


\begin{figure*}[h]
\begin{center}
  \includegraphics[width=\columnwidth]{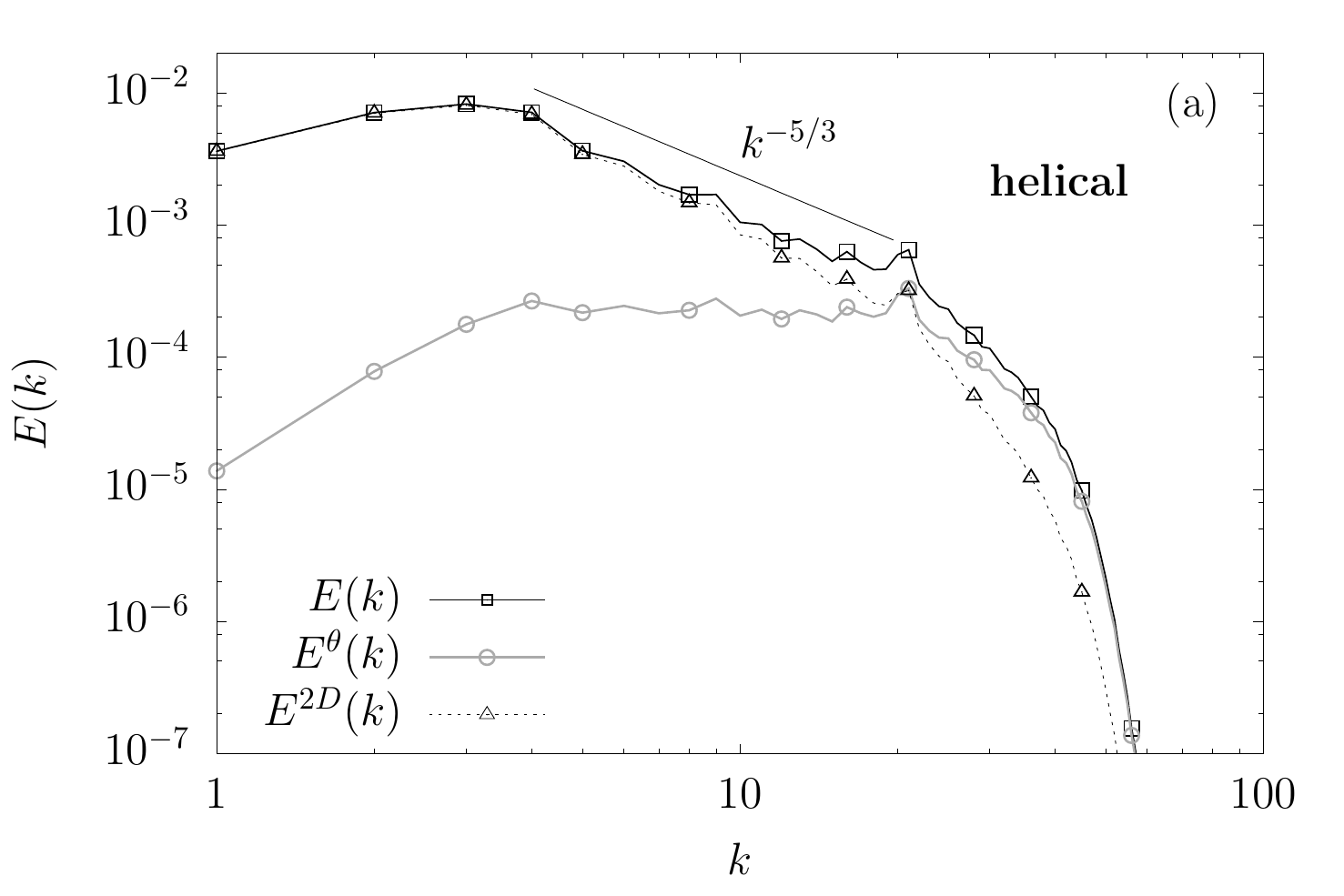}
  \includegraphics[width=\columnwidth]{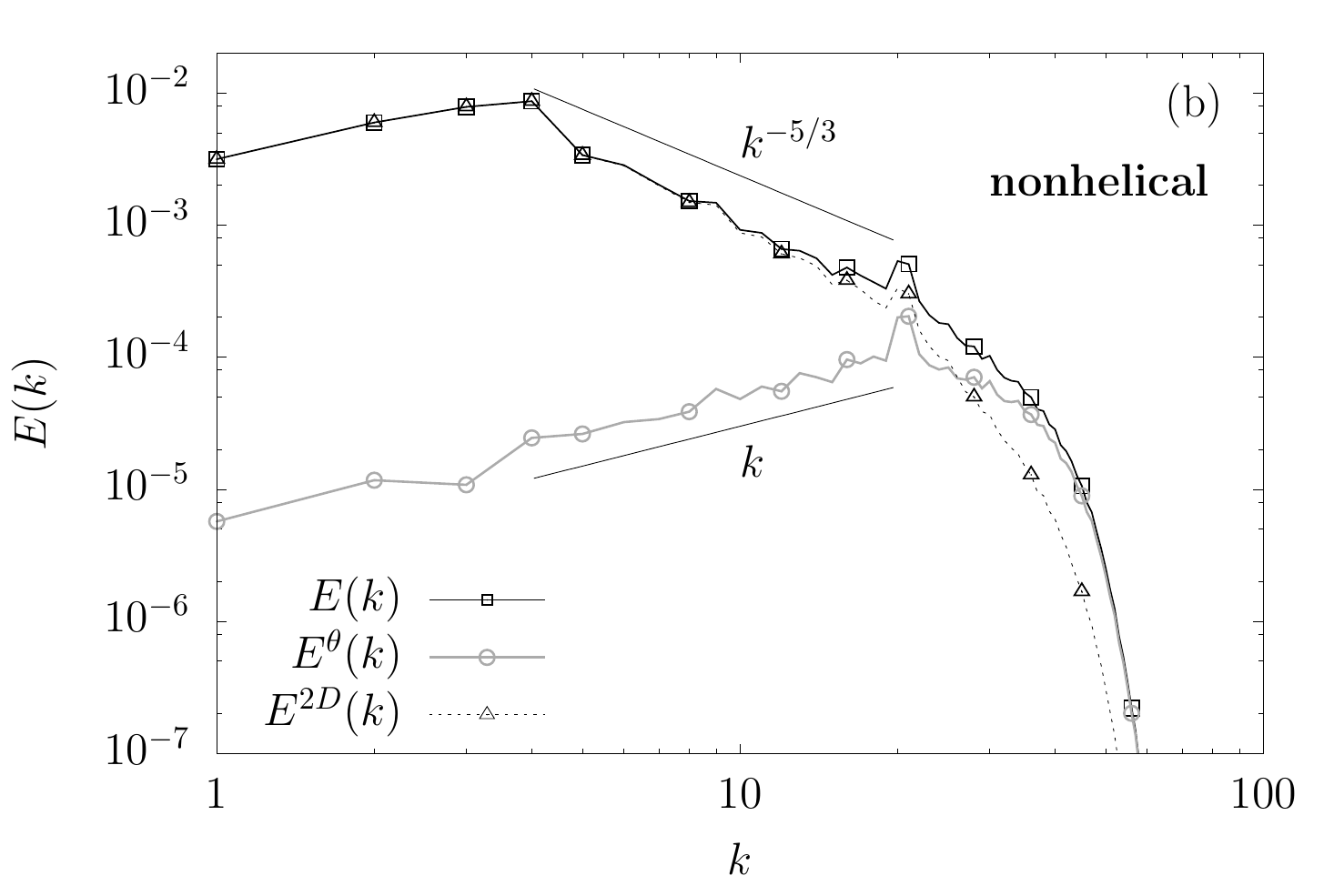}
\end{center}
\caption{Energy spectra for (a) dataset (Hel1), and (b) dataset (NonHel1) at $t/T_f = 25$.}
\label{fig:spectra-fluxes-T60}       
\end{figure*}

\begin{figure*}[h]
\begin{center}
  \includegraphics[width=\columnwidth]{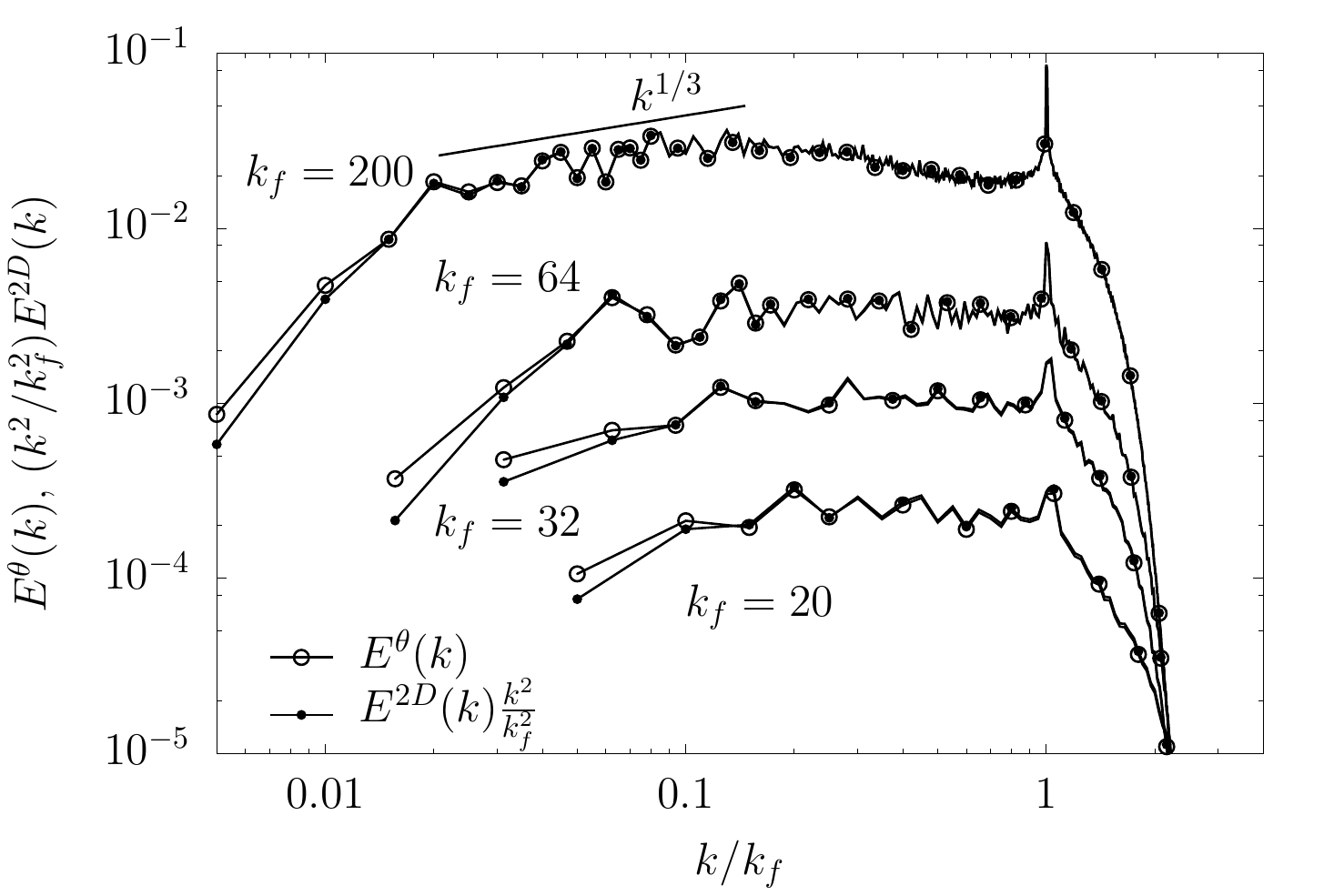}
  \includegraphics[width=\columnwidth]{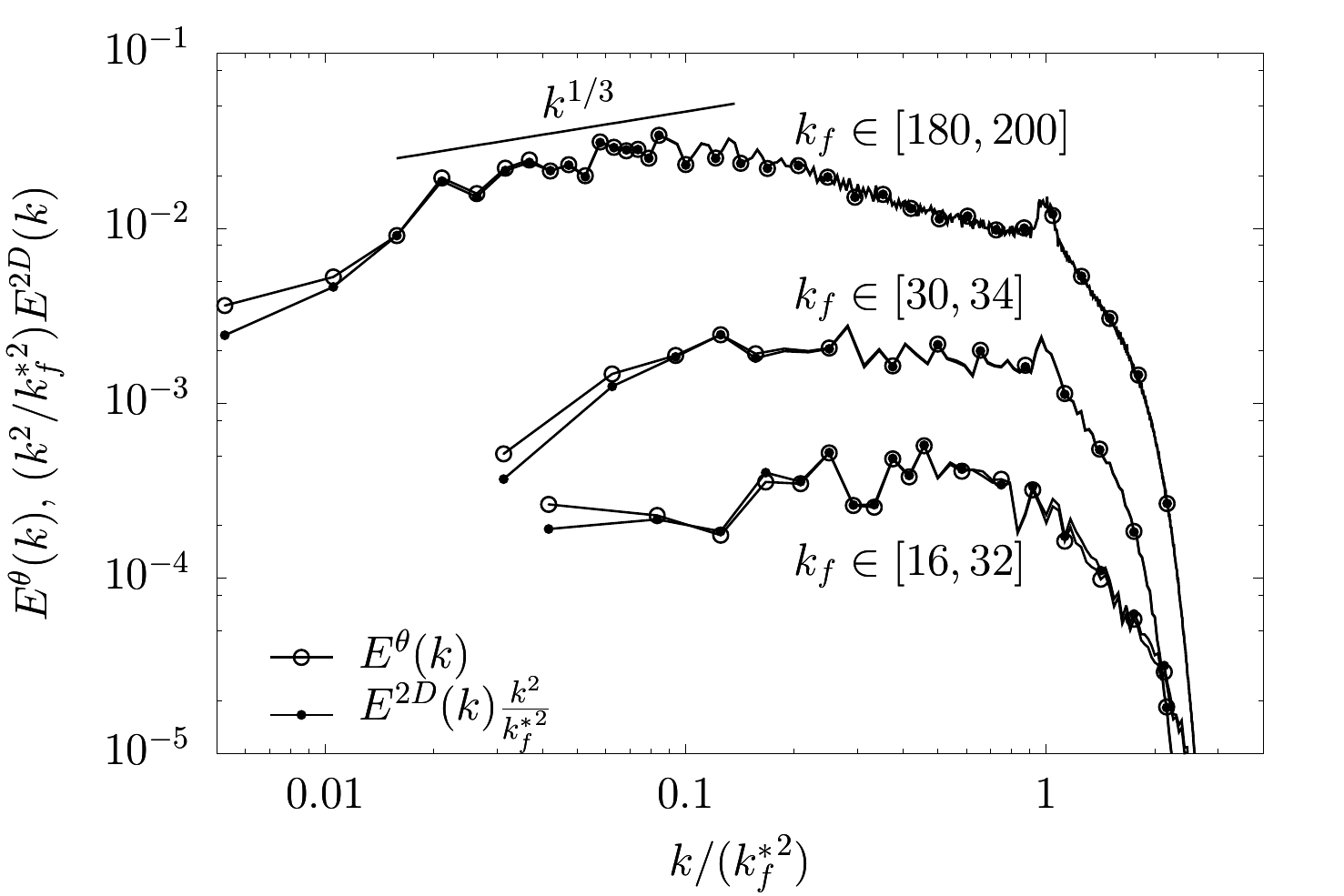}
\end{center}
\caption{\repl{Comparison between $E^\theta(k)$ and $(k^2/k_f^2)E^{\rm 2D}(k)$ for
the different datasets.  
Left: Single-shell forcing, datasets (Hel1), (Hel2), (Hel5) and (Hel6).
Right: Band forcing, runs (Hel3), (Hel4) and (Hel7) with $k_f^* \simeq (k_{\rm min} + k_{\rm max}) / 2$. 
Data corresponds either to time-averages for runs (Hel2) and (Hel3), ensemble average at $t/T_f =
84$ for (Hel1), ensemble average at $t/T_f = 30$ for (Hel6) and (Hel7) or the last snapshot in the simulation for (Hel4) and (Hel5). Spectra corresponding to different datasets are shifted in order to 
facilitate the readability of the figure.}} 
\label{fig:spectra-KF20-KF64-comp}       
\end{figure*}

\begin{figure}[tbp]
  \includegraphics[width=\columnwidth]{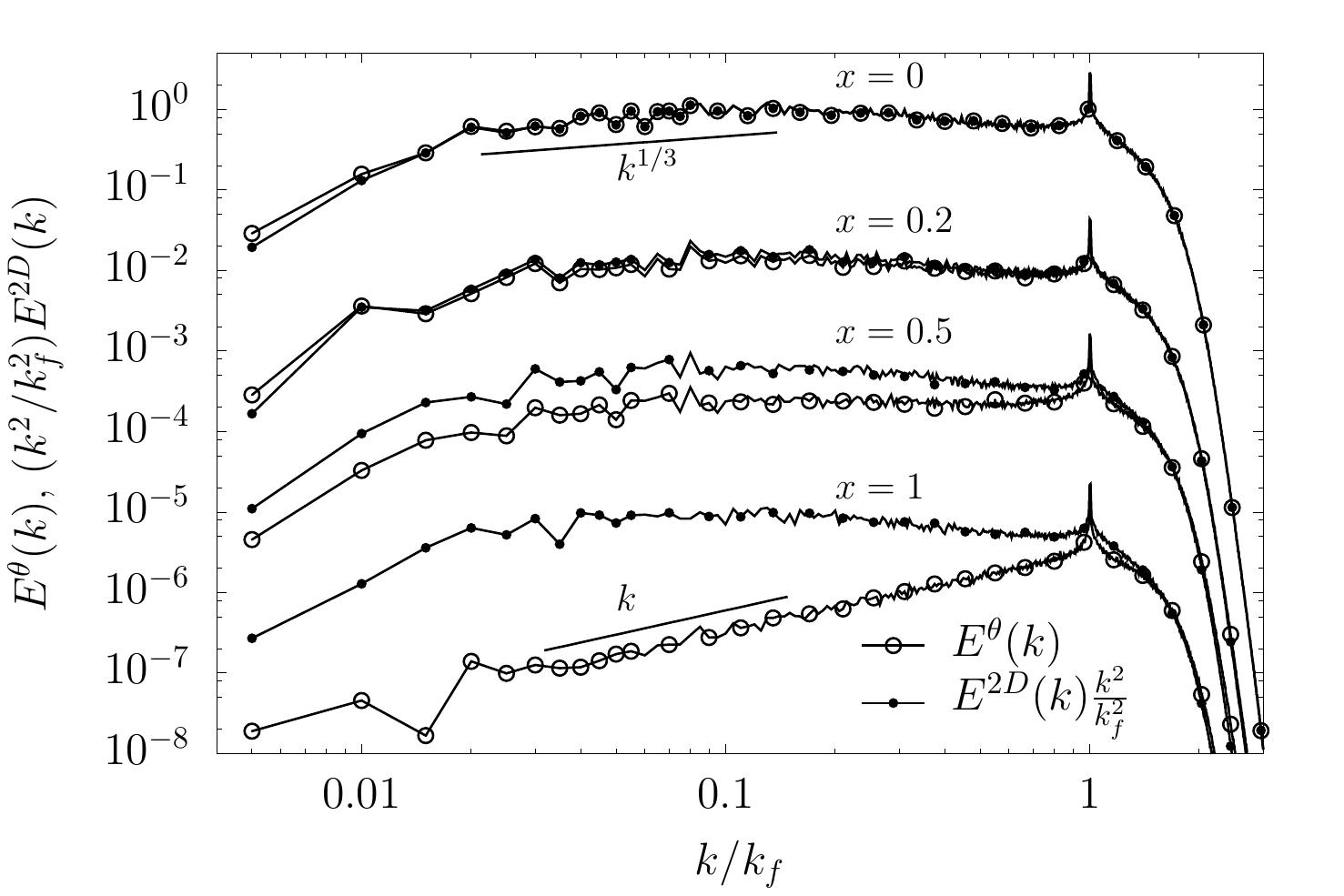}
\caption{
\repl{Comparison between $E^\theta(k)$ and $(k^2/k_f^2)E^{\rm 2D}(k)$ from
simulations forced with different fractions of helicity injection. The database
analysed here consists of runs (Hel6), (FracHel1), (FracHel2) and (FracHel3) 
corresponding respectively to the value of $x=0, \,0.2,\,0.5$ and $1$. 
Data corresponds either to the 
ensemble average at $t/T_f = 30$ for (Hel6) or to the last snapshot in the simulation for the remaining datasets. Spectra corresponding to different datasets are shifted in order to 
facilitate the readability of the figure.}
} 
\label{fig:spectra-mixHel}       
\end{figure}

\begin{figure*}[h]
  \includegraphics[width=\columnwidth]{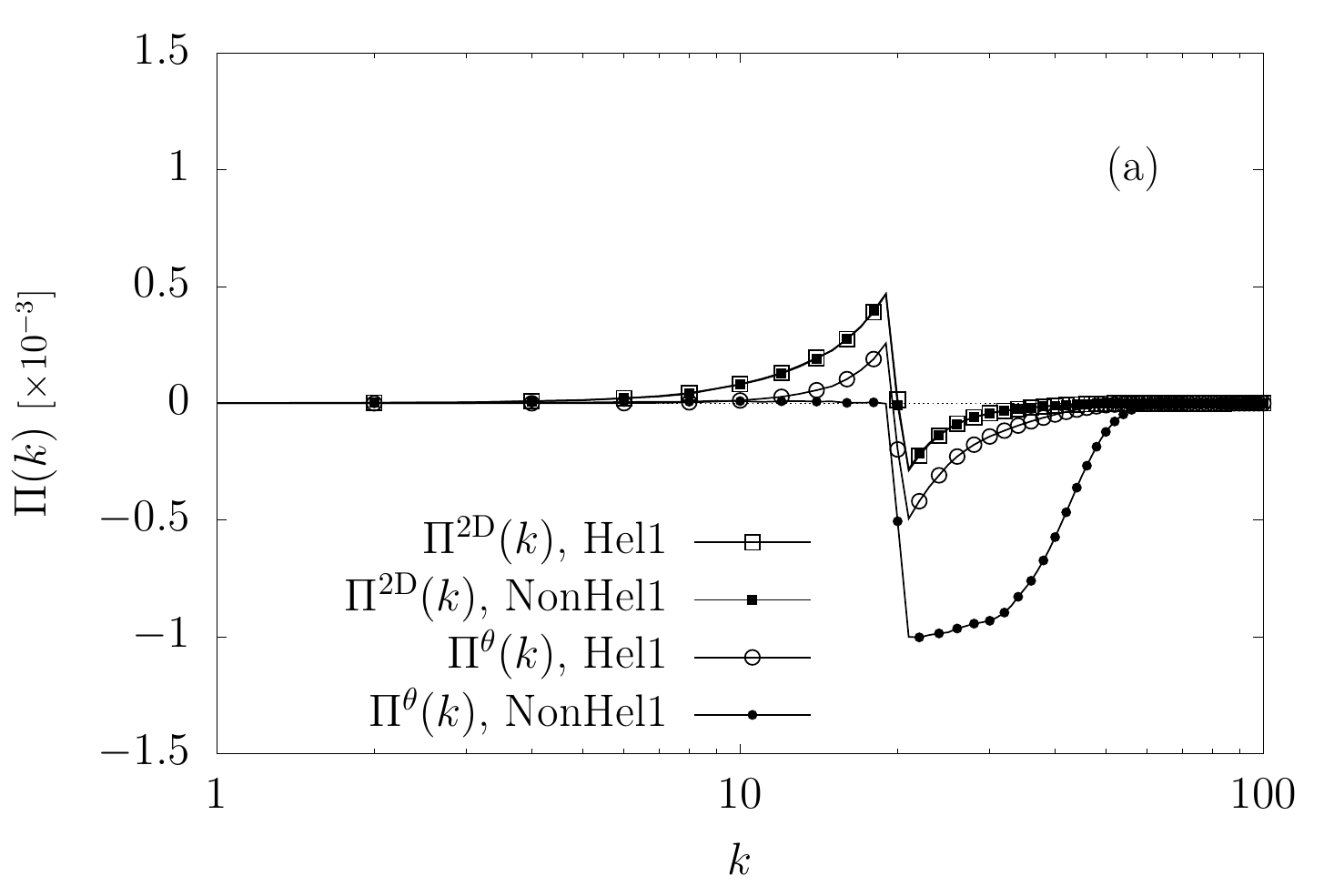}
  \includegraphics[width=\columnwidth]{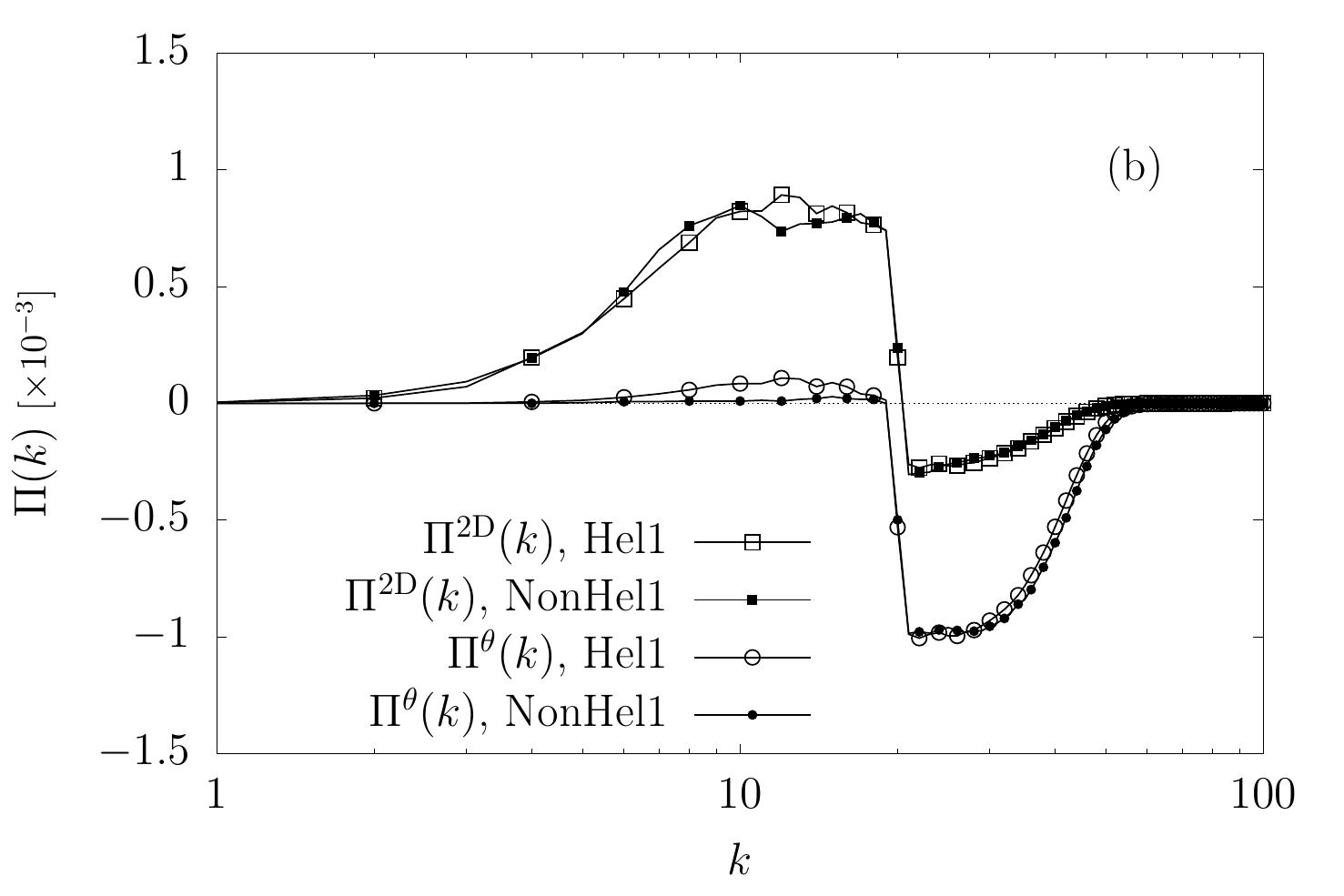}
\caption{
Fluxes of $E^\theta(k)$ and $E^{\rm 2D}(k)$ at early times for datasets (Hel1) and (NonHel1):
(a) $t/T_f=3.4$, (b) $t/T_f=8.4$. 
}
\label{fig:fluxes-T8-T20}       
\end{figure*}

\begin{figure*}[h]
\begin{center}
  \includegraphics[width=\columnwidth]{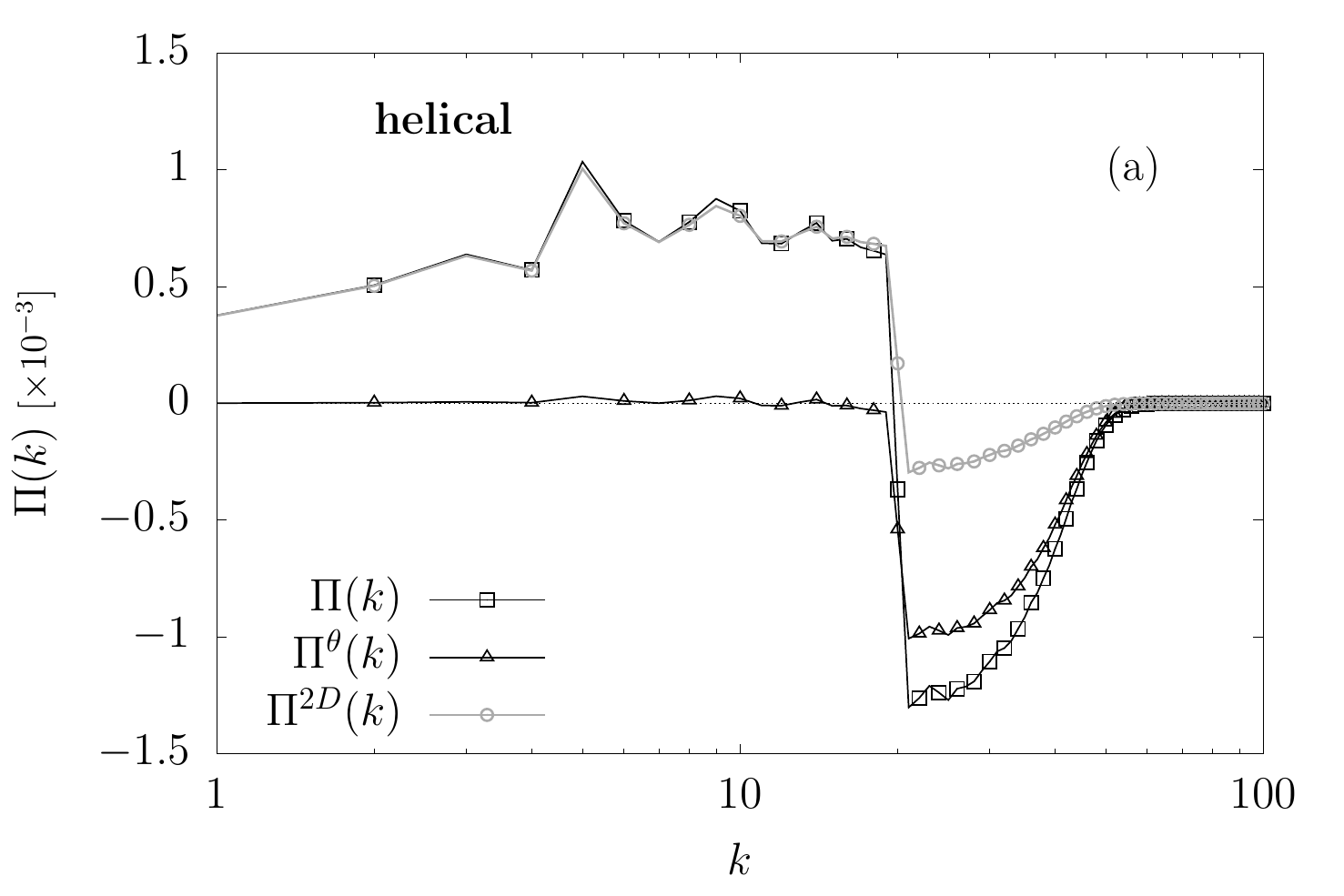}
  \includegraphics[width=\columnwidth]{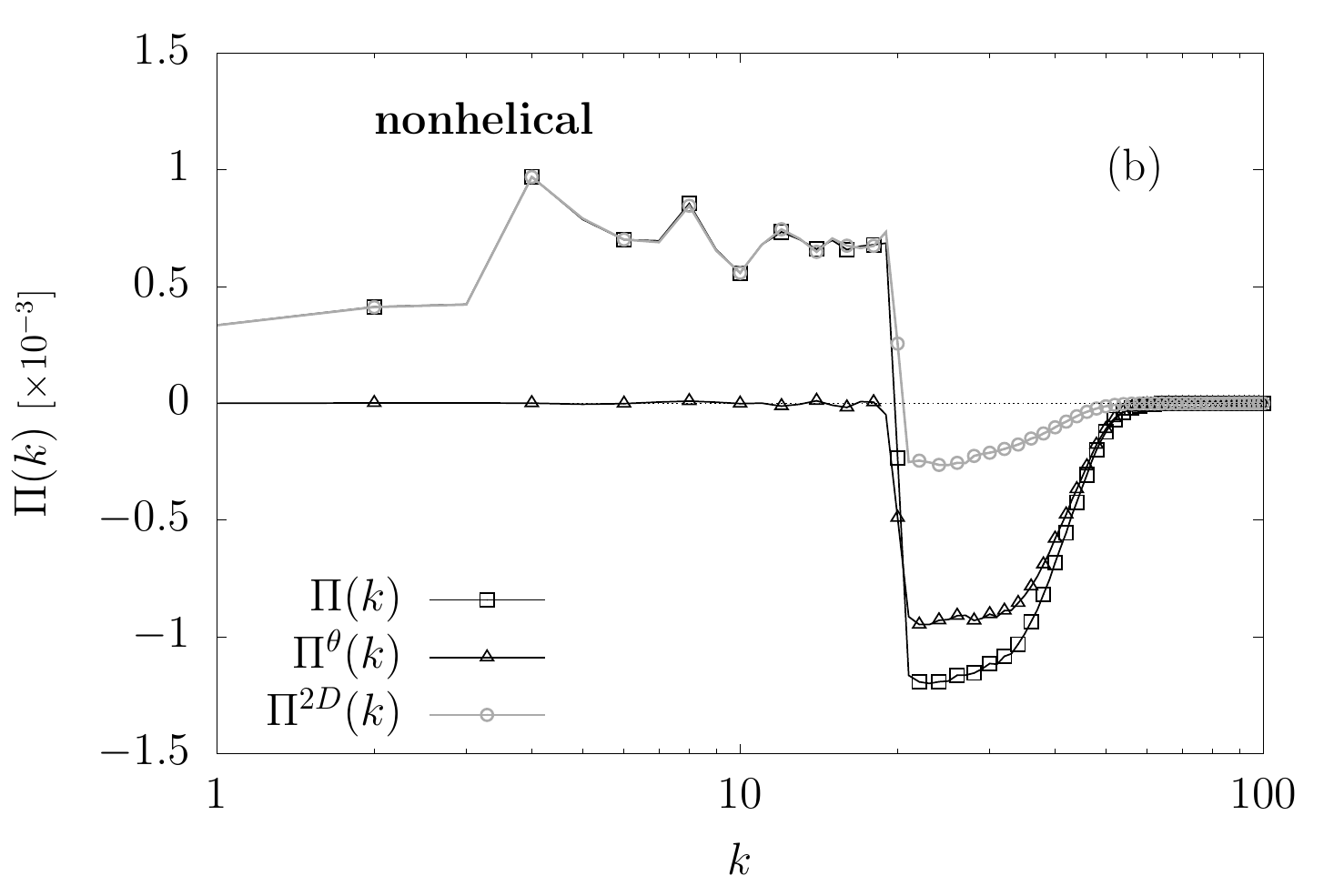}
\end{center}
\caption{Energy fluxes for (a) dataset (Hel1), and (b) dataset (NonHel1) at $t/T_f = 25$.}
\label{fig:fluxes-T60}       
\end{figure*}

\subsection{Spectral scaling}
A further quantitative assessment of the correlation between $\theta$ and $\omega$
can be carried out through a comparison of the energy spectra. 
In particular, we consider the energy spectra of the full velocity field, of
its $\theta$- and 2D-components. 
All three energy spectra obtained from datasets (Hel1) and (NonHel1) 
are shown in 
Fig.~\ref{fig:spectra-fluxes-T60} at $t/T_f = 25$. Two
main phenomenological results can be obtained from this figure. Firstly, as expected, the
2D-component is unaffected by helicity of the forcing, as $\buperp$ displays
inverse-cascade spectra in both cases with a spectral index close to the
predicted $-5/3$ for 2D turbulence \cite{Kraichnan67}. Furthermore, the 2D
spectra are also qualitatively very similar in regions outwith the inertial
range.  Second, the $\theta$-component is clearly sensitive to the helicity of
the forcing, as $E^\theta(k)$ shows the expected 2D absolute equilibrium
scaling, $E^\theta(k) \sim k$, for $k<k_f$ in case of dataset (NonHel1), while no 
such scaling is observed for $E^\theta(k)$ for dataset (Hel1). 
\noindent 
Figure
\ref{fig:spectra-KF20-KF64-comp} provides a comparison between $E^\theta(k)$
and 
$(k/k_f^*)^2E^{\rm 2D}(k)$ for single-shell and band forcing. The left
panel presents the single-shell cases (Hel1), (Hel2) and (Hel5). It can clearly be
seen that the scaling predicted by Eq.~\eqref{eq:spectra-helforce-shell} is in
agreement with the data for all $k$ except the smallest two wavenumbers, 
independent of the
separation between the forcing shell and the smallest resolved wavenumber.
Similar results are obtained for the band-forced case shown in the right panel
for runs (Hel3) and (Hel4), provided $k_f$ is replaced by an `effective forcing wavenumber' 
$k_f^* \simeq (k_{\rm min} + k_{\rm max}) / 2$ for a forcing band
given by the interval $[k_{\rm min},k_{\rm max}]$.
\noindent
In summary, the scaling given by Eq.~\eqref{eq:spectra-helforce-shell} for 
helical forcing also applies for the more general case of band forcing 
to a good approximation. 
For this reason we restrict our attention to single-shell forcing in the 
remainder of this paper, where we focus on the nonuniversal dynamics of 
$\theta$. 
\repl{In Fig.~\ref{fig:spectra-mixHel} the comparison between  $E^\theta(k)$
and $(k/k^*_f )^2 E^{2D}(k)$ for 
fractionally helical forcing is presented. 
In this analysis we have used the following sets of simulations, 
(Hel6) $x=0$, (FracHel1) $x=0.2$, (FracHel2) $x=0.5$ and (FracHel3)
$x=1$ at $2048^2$ collocation points. From
Fig.~\ref{fig:spectra-mixHel} it is visible that upon decreasing the
helicity injected by the forcing, hence upon decreasing the value of $x$, 
the two spectra tend to overlap perfectly following the prediction given in
Eq.~\eqref{eq:spectra-helforce-shell}. Important deviations are already visible
at values of $x$ around $0.5$ while for $x=1$ the scalar component is again
completely passive as expected and it develops an absolute equilibrium 
spectrum at wavenumbers smaller than $k_f$. The results presented in 
Fig.~\ref{fig:spectra-mixHel} clearly show a non-universal nature of 
the passive component, which changes its behaviour depending on the helical properties of the external forcing.}

\subsection{Fluxes and nonuniversal dynamics}
\label{sec:fluxes}
As discussed earlier, while $\buperp$ undergoes an inverse energy cascade,
%
%
$\theta$ appears to become statistically stationary after an initial transient in both
cases (Hel1) and (NonHel1). In case (NonHel1), such observation is in agreement
with the results on passive scalar advection in 2D turbulence, where $\theta$ 
cascades from large to small scales and attains equipartition at scales larger than the 
forcing scale. However,
in case (Hel1) the out-of-plane component is active and forced in a finite band of Fourier
modes (here, a single shell), which implies that the large-scale dynamics of $\theta$ is 
connected to the nonstationary inverse-cascade dynamics of $\bu^{\rm 2D}$. In particular, 
$E^\theta(k,t)$ cannot be statistically stationary at $k < k_f$ according to 
Eq.~\eqref{eq:spectra-helforce-shell}, however, the scaling $E^{\theta}(k,t)\sim k^{1/3}$ results in 
$E^\theta(k,t) \to 0$ for $k \to 0$. The latter implies that the total energy
of $\theta$ in the region $k\leqslant k_f$ is dominated by the contribution at 
the forcing shell
\be
E^\theta(t)_{|k\leqslant k_f} \simeq  \int_{k_0(t)}^{k_f} E^\theta(k,t) dk 
= k_f^{4/3}-k_0(t)^{4/3} \ ,
\ee   
where $k_0(t)$ is the smallest wavenumber at which $\theta$ contains
a significant amount of energy at time $t$. Since $k_0(t) \to 0$ for 
$t \to \infty$, $E^\theta(t)$ indeed tends to a constant for $t \to \infty$, 
even though 
$E^\theta(k,t)$ is not stationary according to 
eq.~\eqref{eq:spectra-helforce-shell}. 
\\

\begin{figure*}[h]
  \includegraphics[width=\columnwidth]{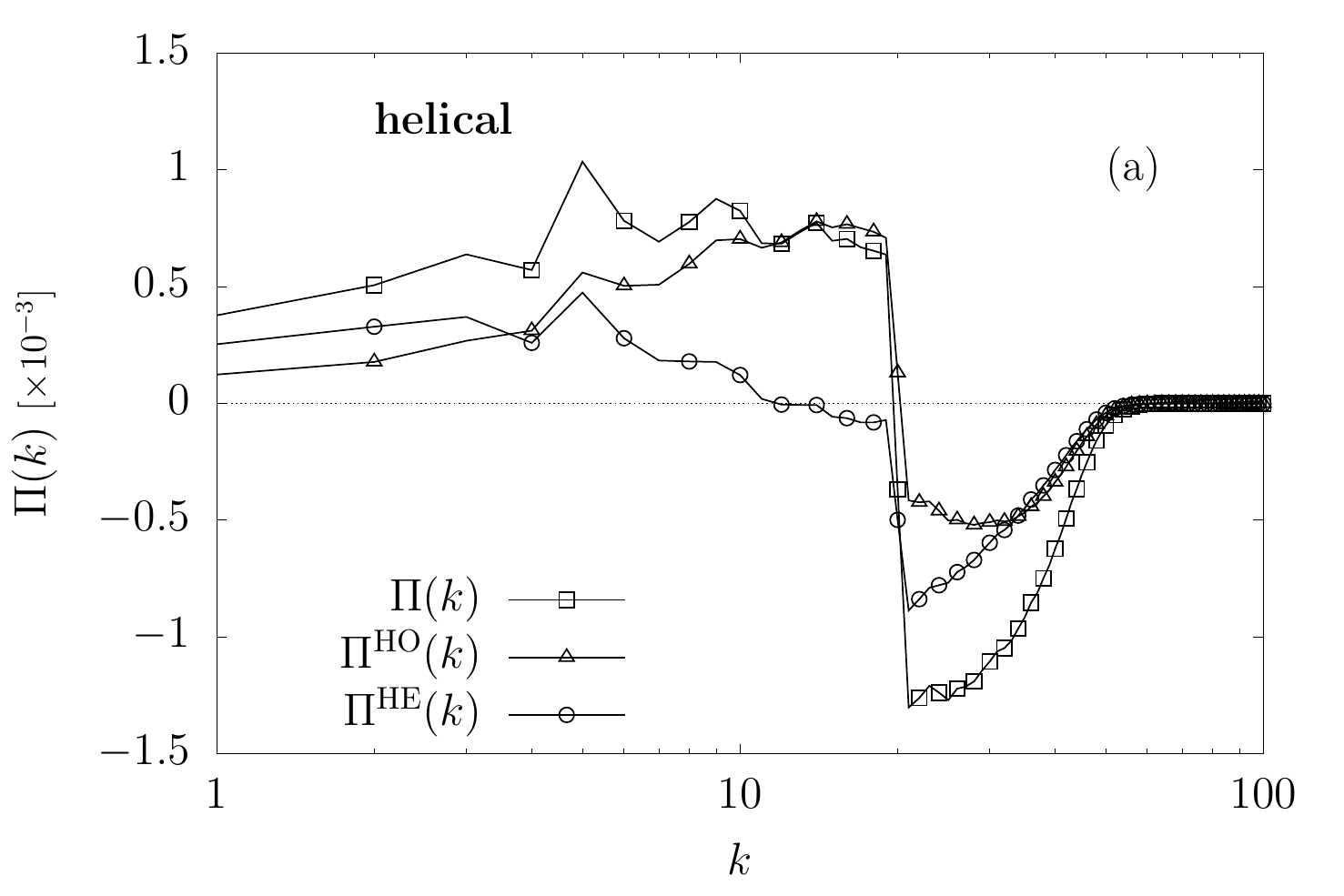}
  \includegraphics[width=\columnwidth]{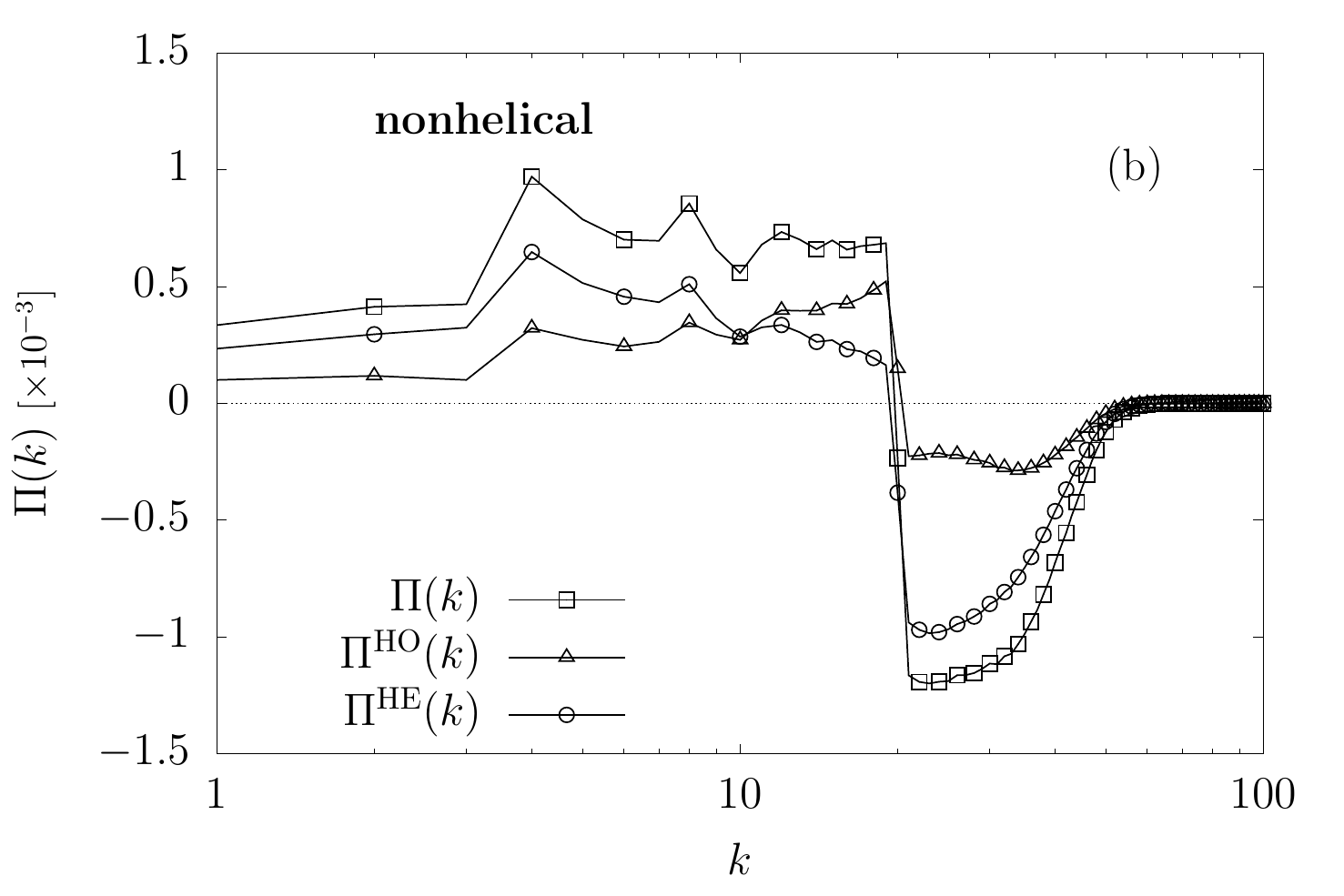}
\caption{
Homo- and heterochiral subfluxes at $t/T_f = 25$. 
(a) $\Pi(k)$ (squares), $\Pi^{\rm HO}(k)$ (triangles) and 
$\Pi^{\rm HE}(k)$ for helical forcing. 
(b) The same for nonhelical forcing. 
}
\label{fig:helfluxes-T60}       
\end{figure*}

\begin{figure*}[h]
  \includegraphics[width=\columnwidth]{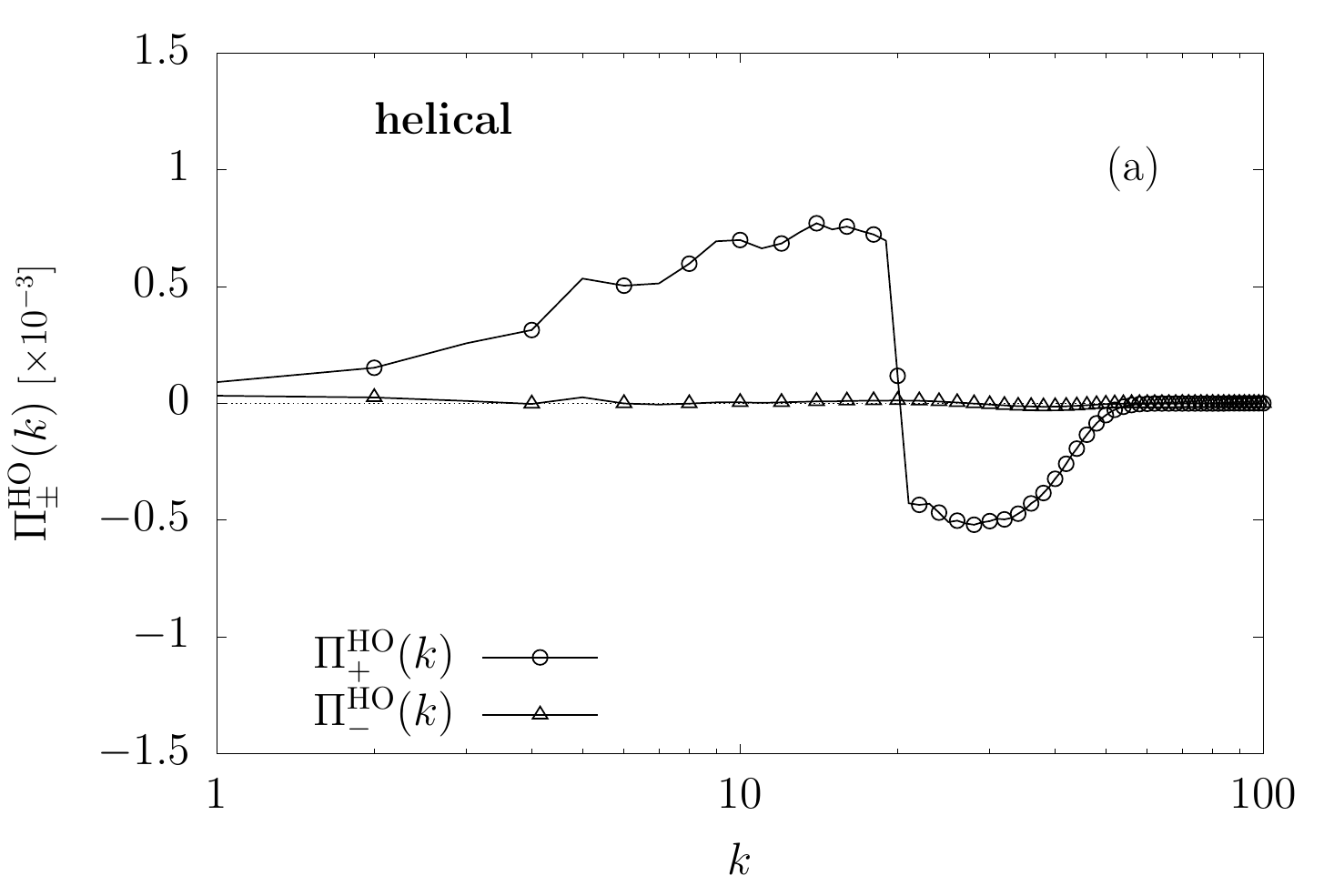}
  \includegraphics[width=\columnwidth]{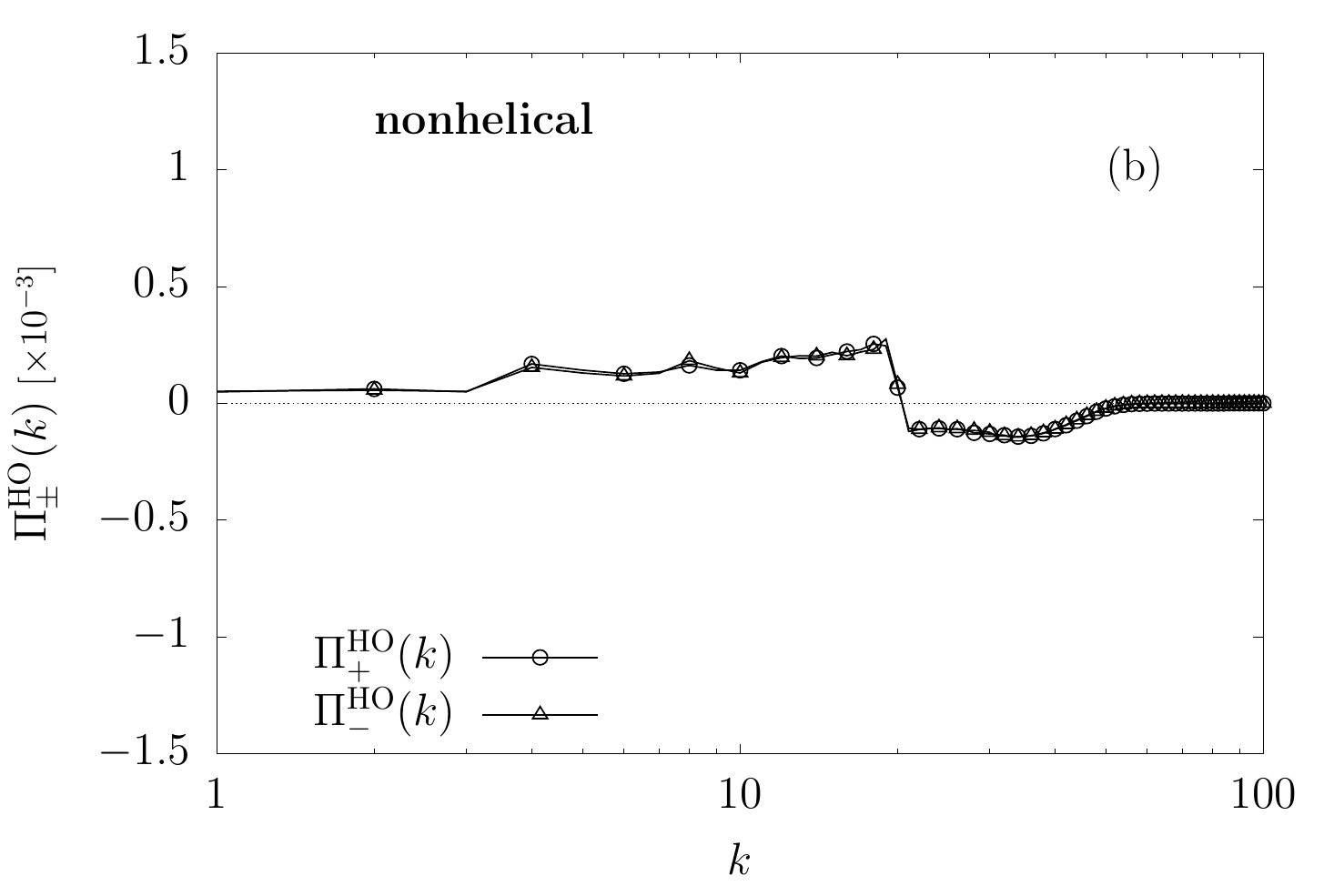}
\caption{
Homochiral subfluxes $\Pi^{\rm HO}_{+}(k)$ (circles) and $\Pi^{\rm HO}_{-}(k)$ (triangles) 
for (a) helical forcing (Hel1) and (b) nonhelical forcing (NonHel1) at  $t/T_f = 25$.
}
\label{fig:homohelfluxes-T60}       
\end{figure*}

\noindent 
In order to understand the matter more clearly, 
it is instructive to consider the energy fluxes 
\begin{align}
\Pi(k)  &= - \sum_{k'=1}^{k}\sum_{|{\bk}|=k'} 
               \hat \bu_\bk \cdot \hspace{-1em} \sum_{\bk+\bp+\bq=0} \hspace{-1em}(i\bk \cdot \hat \bu_\bp) \hat\bu_\bq\ , \\
\Pi^\theta(k)  &=  -\sum_{k'=1}^{k}\sum_{|{\bk}|=k'} 
               \hat\theta_\bk \hspace{-1em} \sum_{\bk+\bp+\bq=0} \hspace{-1em} 
(i\bk \cdot \fbuperp_\bp) \hat\theta_\bq  \ , \\
\Pi^{2D}(k)  &= - \sum_{k'=1}^{k}\sum_{|{\bk}|=k'} 
               \fbuperp_\bk \cdot \hspace{-1em}\sum_{\bk+\bp+\bq=0}\hspace{-1em}
 (i\bk \cdot \fbuperp_\bp) \fbuperp_\bq\ .
\end{align}
While $\Pi^\theta(k)$ must vanish in case (NonHel1) for $k < k_f$, 
the 
correlation between $\theta$ and $\omega$ in case (Hel1) 
should result in a subleading correction which vanishes in the limit $t \to \infty$. 
That is, $\Pi^\theta(k)$ should be nonzero during the transient stage 
for case (Hel1), i.e., before $\Pi^{\rm 2D}(k) = const$ for $k < k_f$ and
$\Pi^\theta(k) = const$ for $k> k_f$ is established. 
The fluxes $\Pi^{\rm
2D}(k,t)$ and $\Pi^\theta(k,t)$ are shown in Fig.~\ref{fig:fluxes-T8-T20} for
two times at $t/T_f=3.4$ and $t/T_f=8.4$. 
As can be
seen, $\Pi^{\rm 2D}(k,t)$ is nearly identical at both times for (Hel1) and (NonHel1),
while the two cases are distinct concerning the behaviour of $\Pi^\theta(k,t)$.
In case (NonHel1), only a direct transfer of $E^\theta(k)$ with a well-established
forward flux is 
present at $t/T_f=3.4$. In particular, there is no inverse
$E^\theta(k)$-flux. In contrast, for case (Hel1), $\Pi^\theta(k,t)$ is depleted
in the direct transfer region $k>k_f$ with 
nonzero inverse $E^\theta(k)$-flux, which decreases with time as can be seen 
by comparing Figs.~\ref{fig:fluxes-T8-T20}(a) and (b). 
Figure \ref{fig:fluxes-T60} presents the fluxes 
$\Pi(k)$, $\Pi^\theta(k)$ and $\Pi^{\rm 2D}(k)$ at $t/T_f = 25$, where 
$E^\theta(t)$ is statistically stationary. 
In the region $k<k_f$ we indeed find $\Pi^\theta(k)
=0$ in both cases, while the entire inverse energy flux is carried
by the 2D component. As can be seen from the figure, the fluxes $\Pi^\theta(k)$
and $\Pi^{\rm 2D}(k)$ are now very similar for the two datasets. The absolute
equilibrium spectral scaling of $E^\theta(k)$ is in accord with $\Pi^\theta(k)
=0$ in the region $k < k_f$ for case (NonHel1), while for the helically forced case
$\theta(k)$ is out of equilibrium. 
In summary, under helical forcing the
out-of-plane component $\theta$ develops a transient inverse transfer, which is
absent in case (NonHel1). \\ 

\noindent
The differences we find in the behaviour of $\theta$ must be caused
by a superabundance of helical velocity field 
modes of one sign close to the forcing shell. Furthermore, 
it is known that helicity cannot play a role in the 2D-dynamics, 
that is, any interaction of helical modes should participate democratically in 
the inverse energy cascade of $\buperp$. 
In this context, it is important to note that 
the transient inverse energy flux of $\theta$ is 
a three-dimensional effect which is intrinsically connected to the 
breaking of mirror symmetry by strongly helical forcing. 
In fact, a fully 3D inverse cascade can be achieved
by breaking mirror symmetry at all scales through projection onto one helical
component \cite{Biferale12,Biferale13a}, and interactions of helical
modes of the same sign also contribute to a subleading 
3D inverse energy transfer in full Navier-Stokes 
dynamics \cite{Alexakis17}. 
Here, we only project the Fourier components of the force onto the
positively helical sector letting the nonlinear interactions generate velocity 
field modes of either sign of helicity. This leads to a dominance
of positively helical modes in the presence of a strong 2D inverse transfer. 
It is therefore of interest to establish which interactions 
of helical modes are now mediating the inverse energy transfer.   
\\ 

\noindent
We first define the energy subfluxes which correspond to 
helical interactions involving 
only velocity modes of a fixed sign of helicity  
\begin{align}
\label{eq:pi-ppp}
\Pi^{\rm HO}_{+}(k)  &=  -\sum_{k'=1}^{k}\sum_{|{\bk}|=k'} 
 \hat \bu^+_\bk \cdot \hspace{-1em} \sum_{\bk+\bp+\bq=0}\hspace{-1em} 
 (i\bk \cdot \hat \bu^+_\bp) \hat\bu^+_\bq\ , \\
\label{eq:pi-mmm}
\Pi^{\rm HO}_{-}(k)  &=  -\sum_{k'=1}^{k}\sum_{|{\bk}|=k'} 
 \hat \bu^-_\bk \cdot \hspace{-1em} \sum_{\bk+\bp+\bq=0}\hspace{-1em} 
 (i\bk \cdot \hat \bu^-_\bp) \hat\bu^-_\bq\ ,
\end{align}
where the label HO stands for homochiral.
As mentioned above, $\Pi^{\rm HO}_{+}(k)$ and $\Pi^{\rm HO}_{-}(k)$ 
contribute to a subleading inverse energy transfer even in 3D. 
Their combined contribution is the total
homochiral flux \cite{Biferale17,Alexakis17}
\be
\label{eq:pi-ho}
\Pi^{\rm HO}(k) = \Pi^{\rm HO}_+(k)+\Pi^{\rm HO}_-(k) \ ,
\ee
while interactions involving velocity modes of oppositely-signed helicity 
are combined to define the heterochiral flux \cite{Biferale17,Alexakis17} 
\be
\label{eq:pi-he}
\Pi^{\rm HE}(k) = \Pi(k)-\Pi^{\rm HO}(k) \ .
\ee
Fully helical forcing
leads to an imbalance between homo- and heterochiral energy fluxes
as shown in Fig.~\ref{fig:helfluxes-T60}. 
For the case (Hel1) we can see that close to the energy
injection scale, where the dynamics is dominated by Fourier modes of 
positive helicity, $\Pi(k)$ is mainly given by $\Pi^{\rm HO}(k)$ with $\Pi^{\rm
HE}(k)$ being negligible, see Fig.~\ref{fig:helfluxes-T60}(a).   
For the case (NonHel1) 
the two contributions are almost identical, as it should be for a 2D 
simulation where helicity does not play any role, see Fig.~\ref{fig:helfluxes-T60}(b).
Owing to the positively helical forcing, in case (Hel1) the
homochiral flux can be expected to be mostly given by its component consisting
of positively helical modes, i.e. $\Pi^{\rm HO}_+(k)$. 
This is indeed the case as can be seen in
Fig.~\ref{fig:homohelfluxes-T60}(a), which presents a comparison between $\Pi^{\rm
HO}_+(k)$ and $\Pi^{\rm HO}_-(k)$ for case (Hel1). Fig.~\ref{fig:homohelfluxes-T60}(b) instead shows that $\Pi^{\rm
HO}_+(k)$ and $\Pi^{\rm HO}_-(k)$ are the same in case (NonHel1). 
\section{Conclusions}
\label{sec:conclusions}
It is known that a sustained 3D inverse energy cascade can be achieved 
by projecting the Navier-Stokes equations onto one helical subspace (homochiral turbulence)
\cite{Biferale12,Biferale13a,Waleffe92}.    
In a homochiral 2D3C flow, the 2D vorticity would be 
fully correlated with  the out-of-plane component 
at all times. As a result,  the out-of-plane component is not passive anymore and develops an inverse cascade similarly to what the
2D velocity field does \cite{Biferale17}. 
Although interesting from a theoretical point of view, such a system 
is somewhat artificial because the projection operation 
must be carried out dynamically in order to remove modes of opposite 
helicity which are generated by nonlinear interactions.
The latter implies that a system 
described by the helically projected Navier-Stokes equations can  be 
studied numerically only.  
Here, we present the application of helical forcing as 
a way of making the out-of-plane component active
while maintaining the full 2D3C Navier-Stokes equations. As such, it is potentially realizable  also  in a  laboratory.
 We show that even in this most realistic case the turbulent transfer for the
 out-of-plane component can be changed through an appropriate forcing, with a sign reversal in the direction of the energy cascade in
 the presence of a strong helical stirring mechanism.
Also here,  nonzero helicity input results in a correlation of the vorticity of the
2D-velocity field with the third component. Such correlation implies that the out-of-plane component can be thought as an active scalar advected by the 2D velocity field.

\noindent 
The correlation between the 2D vorticity
and the third component leads to 
a nonequilibrium large-scale dynamics which is reflected in the scaling of its energy spectrum. While the
energy spectrum of one uncorrelated passive scalar would display absolute
equilibrium scaling, that of the correlated out-of-plane component does not.  
Moreover, we show that the correlation induces a transient 
inverse energy transfer which is mediated 
by homochiral interactions. In other words, a helical input in a 2D3C flow results in a transient 3D contribution to the inverse energy 
transfer, which would be, otherwise, a purely 2D effect in case of no helicity input.  
A similar effect should be present in helically forced flows under 
rapid rotation and in conducting flows in the presence of a strong
mean magnetic field \cite{Alexakis11,buzzicotti2017energy}.

\noindent
\repl{Our study provides further evidence that scalar quantities transported
by a velocity field might develop non-universal transfer properties 
in the presence of a correlation among the injection and the advecting velocity.
Similar effects exist for 2D MHD, where the magnetic potential performs an
inverse cascade while a passive scalar develops a forward transfer.
Furthermore, for surface quasi-geostrophic flows, although both the potential
temperature and a passive dye would  perform a direct cascade, they  have
different scaling laws already at the level of low-order statistical objects.
These significant differences are due to the correlations between the active
scalar input and the advected velocity, as can be seen also by studying the
evolution of Lagrangian trajectories of the two active and passive fields
\cite{celani2004active}.}

\section*{Acknowlegdements}
We acknowledge useful discussions with A. Alexakis and R. Benzi.
The research leading to these results has received funding from the European
Union's Seventh Framework Programme  
(FP7/2007-2013) under grant agreement No. 339032 and from the 
COST Action Programme.

\section*{Authors contribution statement}
All the authors were involved in the preparation of the manuscript.
All the authors have read and approved the final manuscript.
\bibliographystyle{unsrt}
\bibliography{refs,refs_les}

\end{document}